\documentclass[conference]{IEEEtran}
\IEEEoverridecommandlockouts
\usepackage{cite}
\usepackage{amsmath,amssymb,amsfonts}
\usepackage{algorithmic}
\usepackage{graphicx}
\usepackage{url}
\usepackage{tikz}
\usepackage{tabularx}
\usepackage{float}
\usepackage{booktabs}
\usepackage{pgfplots}
\usepackage{subcaption}
\usepackage{makecell}   
\usepackage{textcomp}
\usepackage{xcolor}
\def\BibTeX{{\rm B\kern-.05em{\sc i\kern-.025em b}\kern-.08em
    T\kern-.1667em\lower.7ex\hbox{E}\kern-.125emX}}
\begin{document}

\title{RadarFuseNet: Phase-Weighted Complex-Valued Cross-Attention Fusion for Radar Signal Classification\\

\thanks{The authors would like to thank the Federal Ministry of Research, Technology, and Space (BMFTR) for its support as part of the research program Communication Systems “Souverän. Digital. Vernetzt.”. Joint project 6G-life, project identification number: 16KIS2414}
}

\author{\IEEEauthorblockN{Stefan Hägele\IEEEauthorrefmark{1}, Adam Gorriahn\IEEEauthorrefmark{1}, Philipp Wolters\IEEEauthorrefmark{2} and Eckehard Steinbach\IEEEauthorrefmark{1}}
\IEEEauthorblockA{
\IEEEauthorrefmark{1}\IEEEauthorrefmark{2}\textit{Technical University of Munich, School of Computation, Information and Technology} \\
\IEEEauthorrefmark{1}\textit{Chair of Media Technology} \\
\IEEEauthorrefmark{1}\textit{Munich Institute of Robotics and Machine Intelligence} \\
\IEEEauthorrefmark{2}\textit{Chair of Human-Machine Communication} \\
\{stefan.haegele, adam.gorriahn, philipp.wolters, eckehard.steinbach\}@tum.de}
}

\maketitle

\begin{abstract}
Millimeter-wave (mmWave) radar is a compact sensing technology that is particularly well suited for perception tasks in situations where vision-based sensors are limited, such as under adverse environmental conditions or occlusion.
The complex-valued and nonlinear nature of mmWave radar IQ signals makes complex-valued deep learning a natural choice for extracting relevant information from in-phase and quadrature (IQ) data.
However, progress in IQ-based deep learning is limited by the scarcity of annotated radar IQ datasets.
In this paper, we propose RadarFuseNet, a bidirectional complex-valued cross-attention fusion network with phase-aware weighting inside the attention mechanism, combining IQ and FFT-derived features extracted by two complex-valued CNN feature extractors.
To the best of our knowledge, RadarFuseNet is among the first complex-valued dual-domain fusion frameworks to employ phase-weighted bidirectional cross-attention for radar object classification.
Evaluated on our own custom complex-valued IQ radar dataset of occluded objects, RadarFuseNet achieves classification accuracies of 97.70\% at 64GHz center frequency and 94.70\% at 67GHz center frequency, outperforming all comparison and ablation models.
\end{abstract}


\section{Introduction}
Millimeter-wave (mmWave) radar is an effective sensing modality for various perception tasks, particularly in scenarios where camera- or LiDAR-based sensors are undesirable or impractical.
This advantage is further strengthened by mmWave radar's widespread availability, low cost, and adequate sensing resolution, particularly under challenging conditions such as rain, fog, smoke, or occlusion \cite{adverse1,adverse2,occluded1}, while its distinctive radar signature enables efficient object recognition and classification.
Recent radar deep learning methods commonly operate on image-like representations, such as range-angle and range-Doppler maps, or on sparse point clouds generated using CFAR
\cite{philipp, philipp2}.
Although these representations provide physically meaningful spatial and spectral structure, they can obscure phase relationships present in the original radar measurements.
We therefore investigate direct processing of complex-valued in-phase and quadrature (IQ) signals and their FFT-derived representations.
Previous work in this field has investigated the classification of various objects (occluded by light materials) with the goal of enhancing indoor perception with mmWave radar through direct processing of complex-valued IQ signals \cite{occnet, accor}. However, the task still leaves room for improvement in fully exploiting the complementary representation of complex-valued radar signals.
To improve feature representation while directly processing complex-valued radar signals, we propose RadarFuseNet, a dual-branch fusion network that combines raw IQ and FFT-derived radar features using a phase-weighted complex-valued cross-attention mechanism for computing similarity scores. For this purpose, we use two complex-valued CNNs to extract separate features from the IQ signal and its FFT-transformed counterpart, which are then fused by the proposed complex-valued cross-attention mechanism.
This phase-weighted cross-attention adaptation leverages two complementary complex-valued feature spaces to extract and fuse richer representations. It keeps all possible computations in the complex domain, considering phase information throughout the processing steps.
Although the theoretical information content is identical in both signal representations, our goal is to optimize the feature space and thus feature availability, thereby providing the final model with a more effective learning foundation.
Our main contributions are summarized as follows:
\begin{itemize}
    \item A dual-branch architecture fusing two complex-valued CNNs via bidirectional cross-attention tailored to complex-valued radar signals.
    \item A phase-weighted complex-valued cross-attention formulation for extracting similarity scores directly in the complex domain.
    \item Consistent classification performance across two frequency bands, validated through comprehensive comparisons and ablations.
\end{itemize}
The remainder of this paper is organized as follows. Section~\ref{related_work} reviews related work in recent mmWave radar research and radar-based deep learning. Section~\ref{methodology} presents the proposed methodology, including signal preprocessing and model architecture. Section~\ref{results} reports the experimental results and evaluation, while Section~\ref{conclusion} concludes the paper and outlines limitations as well as directions for future research.
\section{Related Work}
\label{related_work}
Millimeter-wave (mmWave) radar became popular in recent years, driven by affordable, high-resolution hardware and applications spanning autonomous driving, object tracking, gesture recognition, and human activity recognition. Its robustness under adverse environmental conditions further makes it a valuable complement to existing perception approaches, particularly in the automotive domain.
Most deep learning-based approaches fuse RGB-D images with radar or LiDAR data to enhance robustness and depth estimation, with recent methods increasingly incorporating attention-based models alongside CNNs to further improve fusion quality \cite{philipp, philipp2, philipp3, automotive1, automotive2, automotive3}.
However, these fusion approaches typically operate on radar heatmap images (range–angle or range–Doppler) or point clouds rather than the complex-valued analog-to-digital converted (ADC) signal, which retains both amplitude and phase information otherwise lost during image- or point-cloud conversion.
This limitation stems from the scarcity of public datasets with raw complex-valued data and a preference for more interpretable, human-observable representations, a pattern also seen in gesture recognition and human activity datasets \cite{gesture1, human1}.
Similarly, beyond traditional micro-Doppler-based deep learning approaches, recent research has shown attention-based models further enhance robustness and accuracy \cite{gesture2,gesture3,human2}.
Applying attention mechanisms to radar IQ signals introduces a new challenge, as the standard attention formulation is defined in the real domain, and its softmax function requires real-valued input.
This issue is commonly addressed by workarounds such as concatenating the real and imaginary parts as real-valued input, by using a naive complex attention formulation $A(z) = A(\Re\{z\}) + \mathrm{j}A(\Im\{z\})$, or by relying on magnitude-only calculations, mostly ignoring the signal phase and commonly applied in the context of self-attention, with cross-attention left largely unconsidered~\cite{smith2023complexvaluedneuralnetworksdatadriven,eilers2023cplxtrafo,huang2026holoTR, yang2020cplxtransformer}.
While feature fusion typically combines fundamentally different sensor modalities to enhance perception, limited research has explored fusing different representations of the same mmWave radar signal to improve feature extractability rather than informational content.
Although recent research has investigated GNSS signals using time–frequency domain fusion, mmWave radar as well as complex-valued attention fusion was not the subject of evaluation in these studies~\cite{GNSS}.
Due to the scarcity of public datasets containing raw complex-valued radar IQ signals, we rely on two previously recorded custom sub-datasets from \cite{occnet} and \cite{accor}, targeting occluded-object classification across two 4\,GHz-wide bands centered at 64\,GHz and 67\,GHz.
\cite{fabian} and \cite{smcnet} introduced a complex-valued CNN for processing complex-valued radar inputs, which is adopted as the feature extractor in our proposed model.
Both sub-datasets were recorded using a high-resolution mmWave FMCW MIMO imaging radar sensor from Vayyar with an L-shaped $20\times20$ antenna configuration, operating within an RF frequency range of 62–69\,GHz \cite{vayyar}.
%
%
\section{Methodology}
\label{methodology}
\subsection{Signal Preprocessing}
\begin{figure}
    \centering
    \includegraphics[width=0.49\textwidth]{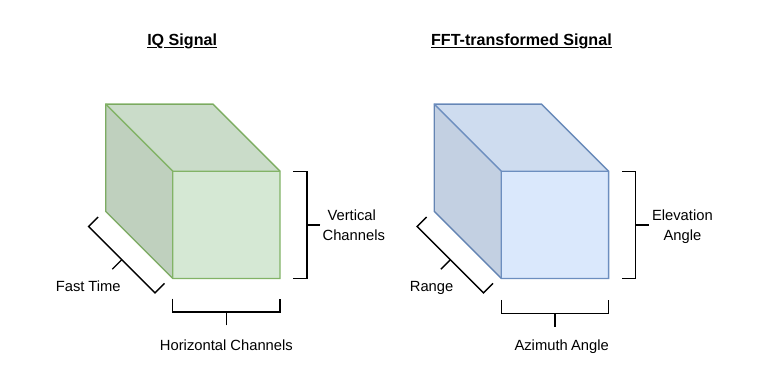}
    \caption{Dimensions of the raw IQ signal (left) and its FFT-transformed counterpart (right).}
    \label{datacube}
\end{figure}

With its 20$\times$20 L-shaped antenna configuration, the radar sensor provides a 400-channel sample $\boldsymbol{s}$ per measurement, with each channel sampled at 100 time points.
The dimensions of $\boldsymbol{s}$ are shown in \eqref{s}.
Each sample contains a single FMCW chirp, as Doppler information is not required and hence omitted due to the static nature of the targets.
In general, the signal can be represented as a 3D data cube, illustrated in Fig.~\ref{datacube}, along with its respective dimensional interpretation.
\begin{equation}
\boldsymbol{s} \in \mathbb{C}^{20\times20\times100}
\label{s}
\end{equation}
To fuse time- and frequency domain data, spectral and raw IQ versions of the signal are required. A 3D fast Fourier transform (FFT) is applied to the signal cube $\boldsymbol{s}$ and provides the range-, azimuth- and elevation spectra $\boldsymbol{S}$ of the signal.
This process is formalized in \eqref{fft2}, where $n$ denotes the fast time dimension and $x$ and $y$ represent the spatial dimensions:
\begin{equation}
\boldsymbol{S}(l,m,k) = \sum_{x=0}^{X-1} \sum_{y=0}^{Y-1} \sum_{n=0}^{N-1} \boldsymbol{s}(x,y,n) e^{-j2\pi\left(\frac{l x}{X}+\frac{m y}{Y}+\frac{k n}{N}\right)}
\label{fft2}
\end{equation}
$l$, $m$, and $k$ are the corresponding transformed indices, representing the range and angle bins in the respective directions, with the same size as $\boldsymbol{s}$.
To meet the input requirements of the complex-valued CNN, both signals $\boldsymbol{s}$ and $\boldsymbol{S}$ are reshaped into 2D representations $\boldsymbol{x}$ and $\boldsymbol{X}$ by flattening along the channel dimension, such that
\begin{equation}
\boldsymbol{x,X} \in \mathbb{C}^{(20\cdot20)\times100}.
\label{2dfft}
\end{equation}
This results in two signal types ready to be used as CNN inputs, $\boldsymbol{x}$ representing the time-domain IQ signal and $\boldsymbol{X}$ its FFT-transformed counterpart, which includes the azimuth and elevation angle profiles as well as the range profile in the complex domain.

\subsection{Model Architecture}
\begin{figure}
    \centering
    \includegraphics[width=0.45\textwidth]{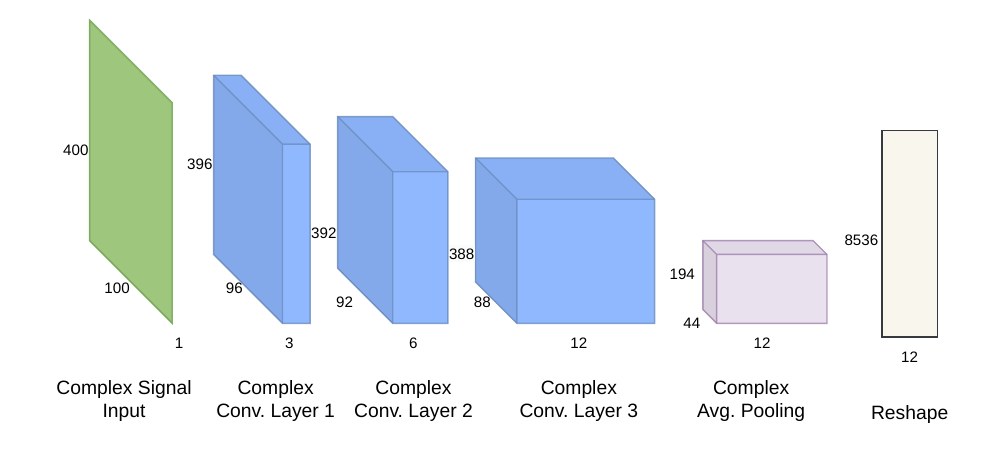}
    \caption{Design of the single-branch complex-valued CNN feature extractor with complex-valued phasor as input.}
    \label{cnn}
\end{figure}

Two distinct complex-valued CNNs are used to extract features in RadarFuseNet. One extracts features from the raw IQ signal $\boldsymbol{x}$, while the other extracts features from its FFT-transformed counterpart $\boldsymbol{X}$.
All operations within these feature-extracting CNNs are performed in the complex domain, including convolutions, pooling, batch normalization, and activation functions \cite{fabian}.
The complex-valued convolution and complex-valued tanh activation function are given in \eqref{conv} and \eqref{tanh}, respectively.
Activation, batch normalization \eqref{batch}, and pooling are applied naively, performed separately on the real and imaginary components, which simplifies backpropagation compared to their analytic counterparts~\cite{naive-complex}.
In general, complex-valued convolutional layers are advantageous, as they enable direct processing and feature extraction from the complete signal without separating its real and imaginary components.
This allows features to be derived from the intact signal phasor rather than from a split and concatenated representation.
For kernels $k = a+\mathrm{j}\,b \in \mathbb{C}$ and input $x = c+\mathrm{j}\,d \in \mathbb{C}$, the convolution is given by
\begin{equation}
    k \cdot x = c \cdot a - d \cdot b + \mathrm{j}\,(c \cdot b + d \cdot a),
    \label{conv}
\end{equation}
and for batch normalization ($BN$) of $z \in \mathbb{C}$
\begin{equation}
    \mathbb{C}BN(z) = BN(\Re\{z\}) + \mathrm{j}\,BN(\Im\{z\}).
    \label{batch}
\end{equation}
The complex-valued tanh activation function is defined as
\begin{equation}
\mathbb{C}\tanh(z) = \tanh(\mathrm{\Re}(z)) + \mathrm{j}\,\tanh(\mathrm{\Im}(z))
    \label{tanh}
\end{equation}
with
\begin{equation}
\tanh(a) = \frac{e^{a} - e^{-a}}{e^{a} + e^{-a}},
\end{equation}
as it outperforms the complex-valued ReLU counterpart here.
\begin{figure}
    \centering
    \includegraphics[width=0.49\textwidth]{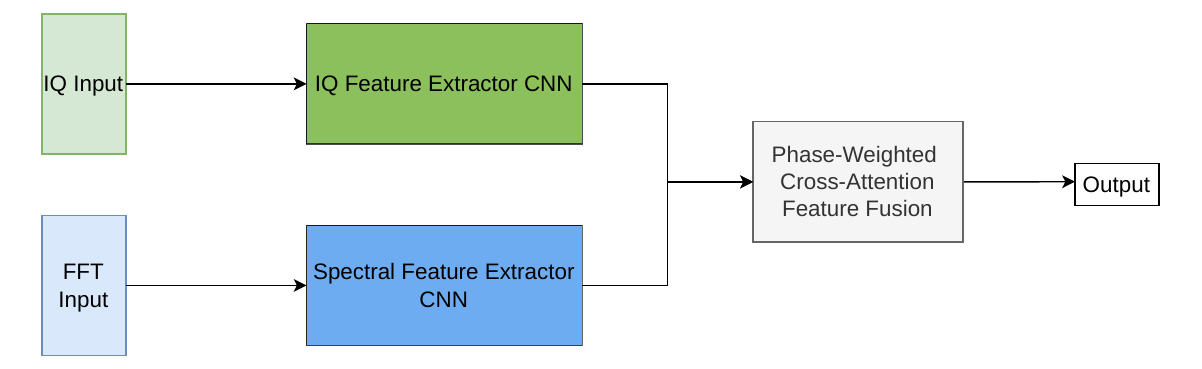}
    \caption{Model architecture including two complex-valued CNNs for IQ- ($\boldsymbol{z}_x$) and FFT ($\boldsymbol{z}_X$) feature extraction, and the multi-head phase-weighted cross-attention mechanism to fuse the extracted features.}
    \label{model-fig}
\end{figure}
The overall architecture is shown in Fig.~\ref{model-fig}.
Both CNN branches share the same structure.
The feature extractors are depicted in Fig.~\ref{cnn}, realized with three complex-valued layers and complex-valued average pooling. Each complex convolutional layer uses a kernel size of $5\times5$, followed by a final complex average pooling with a kernel size of $2\times2$. The feature map channels increase from $1\rightarrow3\rightarrow6\rightarrow12$, resulting in a global feature map of size $8536\times12$ after flattening and concatenating the single feature maps.
For feature fusion, the complex-valued features \(\boldsymbol{z}_x\) (IQ features) and \(\boldsymbol{z}_X\) (FFT features) are projected onto query, key, and value representations and fused by a complex-valued cross-attention module operating directly on these complex representations.
\begin{equation}
\begin{gathered}
    Q_{x/X} = \boldsymbol{z}_{x/X} W_Q, \quad K_{x/X} = \boldsymbol{z}_{x/X} W_K, \\
    V_{x/X} = \boldsymbol{z}_{x/X} W_V,
\end{gathered}
\label{qkv-proj}
\end{equation}
with \(W_Q, W_K, W_V \in \mathbb{C}^{d_{\mathrm{raw}} \times d}\).
Here, \(m = n = 12\) tokens with raw feature dimension \(d_{\mathrm{raw}} = 8536\) are projected into a shared model dimension \(d = 64\), which is then split evenly across \(h = 8\) attention heads, yielding a per-head dimension of \(d_k = d_v = d/h = 8\).
The scaled attention logits \(S\) are then calculated as
\begin{equation}
    S = \frac{QK^{H}}{\sqrt{d_k}} \quad \text{with} \quad S \in \mathbb{C}^{m \times n}.
\end{equation}
To attend the inputs to each other, our proposed phase-weighted complex-valued cross-attention $A_{\mathbb{C}}$ is given by
\begin{equation}
    A_{\mathbb{C}} = \mathrm{softmax}\big(|S| + \lambda \cos(\Delta\phi)\big)V
\end{equation}
with
\begin{equation}
    \Delta\phi = \arg(S) \in [-\pi, \pi]^{m \times n},
\end{equation}
where $\lambda$ serves as a tunable hyperparameter with $\lambda \in [0, 1]$ and $\Delta\phi$ is the phase of the complex-valued similarity score $S$, which is usually discarded when using real-valued representations.
In our implementation, the observed magnitude satisfies $|S|\in[0,0.4]$.
Therefore, $\lambda$ is tuned empirically to control the relative influence of the phase term.
The attention weights are computed using a phase-weighted softmax function, adding relative phase alignment into the magnitude-based similarity while keeping the attention logits real-valued and compatible with softmax.
The term $\cos(\Delta\phi)$ is used because it provides a bounded measure of phase alignment that can be added directly to the magnitude of the attention logits.
Finally, the full phase-weighted cross-attention formula is given by
\begin{equation}
    A_{\mathbb{C}}(Q,K,V)
    =
    \mathrm{softmax}\left(
        \left|\frac{QK^H}{\sqrt{d_k}}\right|
        + \lambda \cos(\Delta\phi)
    \right)V.
\end{equation}
This is equal to the standard cross-attention formula, augmented with a phase-weighting term inside the softmax function.
This procedure is applied to both feature streams in the same manner to ensure positional coherence. IQ and FFT representations are fused bidirectionally.
Further, positional encoding is included via a shared set of learnable complex-valued embeddings, added to the projected tokens of both the IQ and FFT streams prior to cross-attention.
The proposed cross-attention is then applied as formulated in \eqref{combined}.
For feature inputs $\boldsymbol{z}_x$ and $\boldsymbol{z}_X$, the following operations are performed:
\begin{equation}
    A_1 = A_{\mathbb{C}}(Q_x,K_X,V_X)
    \qquad
    A_2 = A_{\mathbb{C}}(Q_X,K_x,V_x)
    \label{bidirectional-attention}
\end{equation}
\begin{equation}
    A = \text{concat}(A_1, A_2)
    \label{combined}
\end{equation}
As the task is classification, a single fully connected layer followed by a cross-entropy loss function is applied to obtain the final prediction.
The cross-entropy loss is defined as
\begin{equation}
    H(\boldsymbol{\hat{\xi}},\boldsymbol{\xi}) = - \sum_{j=1}^{C} \xi_{j} \log(\hat{\xi}_{j}),
    \label{crossentropy}
\end{equation}
where $\boldsymbol{\hat{\xi}}$ and $\boldsymbol{\xi}$ are the predicted and target distributions, respectively. $C$ represents the number of object classes.
\section{Results and Ablation Study}
\label{results}
\begin{figure}[htbp]
    \centering
    \begin{subfigure}[b]{0.14\textwidth}
        \centering
        \includegraphics[width=\textwidth]{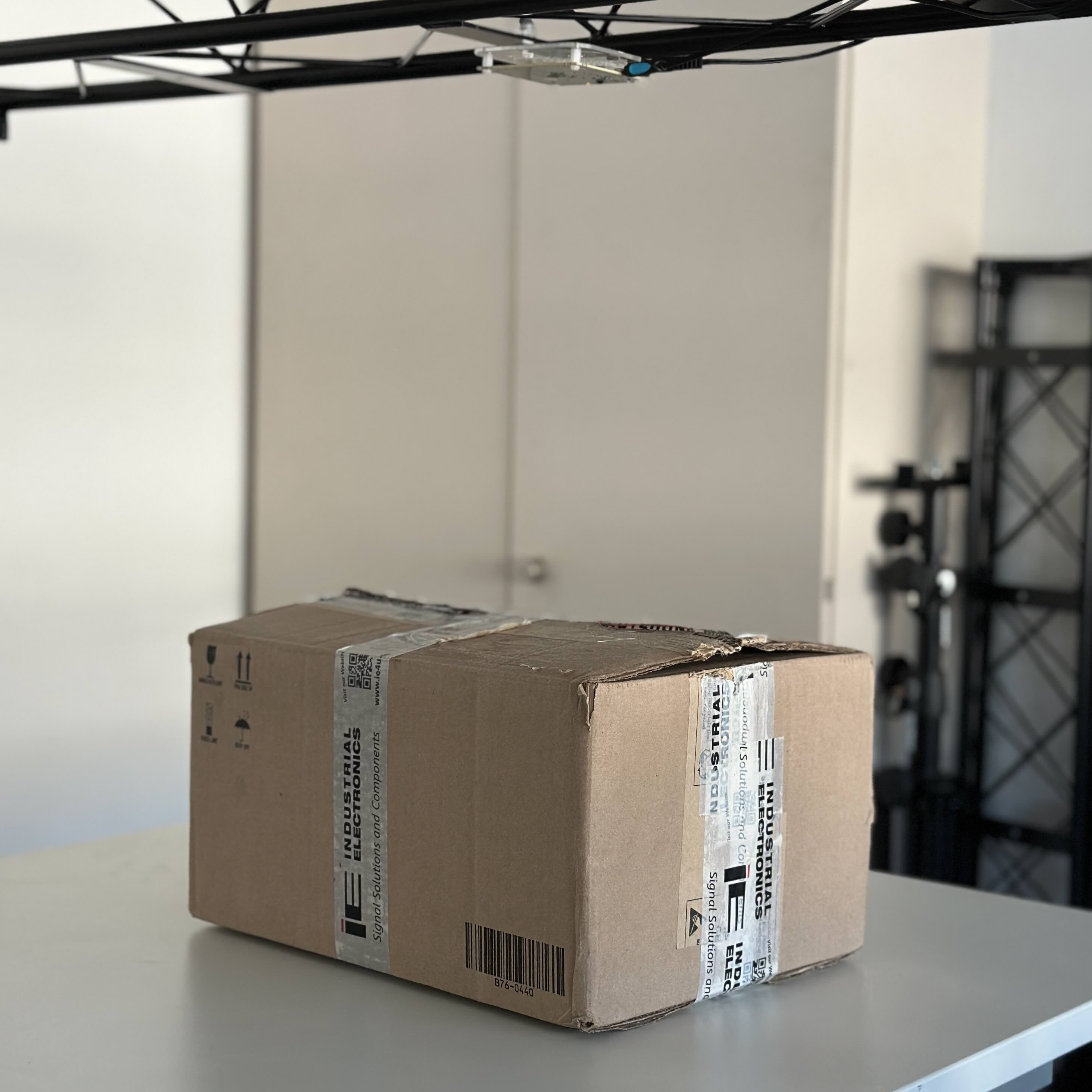}
        \label{fig:setup_a}
    \end{subfigure}
    \hspace{0.5em}
    \begin{subfigure}[b]{0.14\textwidth}
        \centering
        \includegraphics[width=\textwidth]{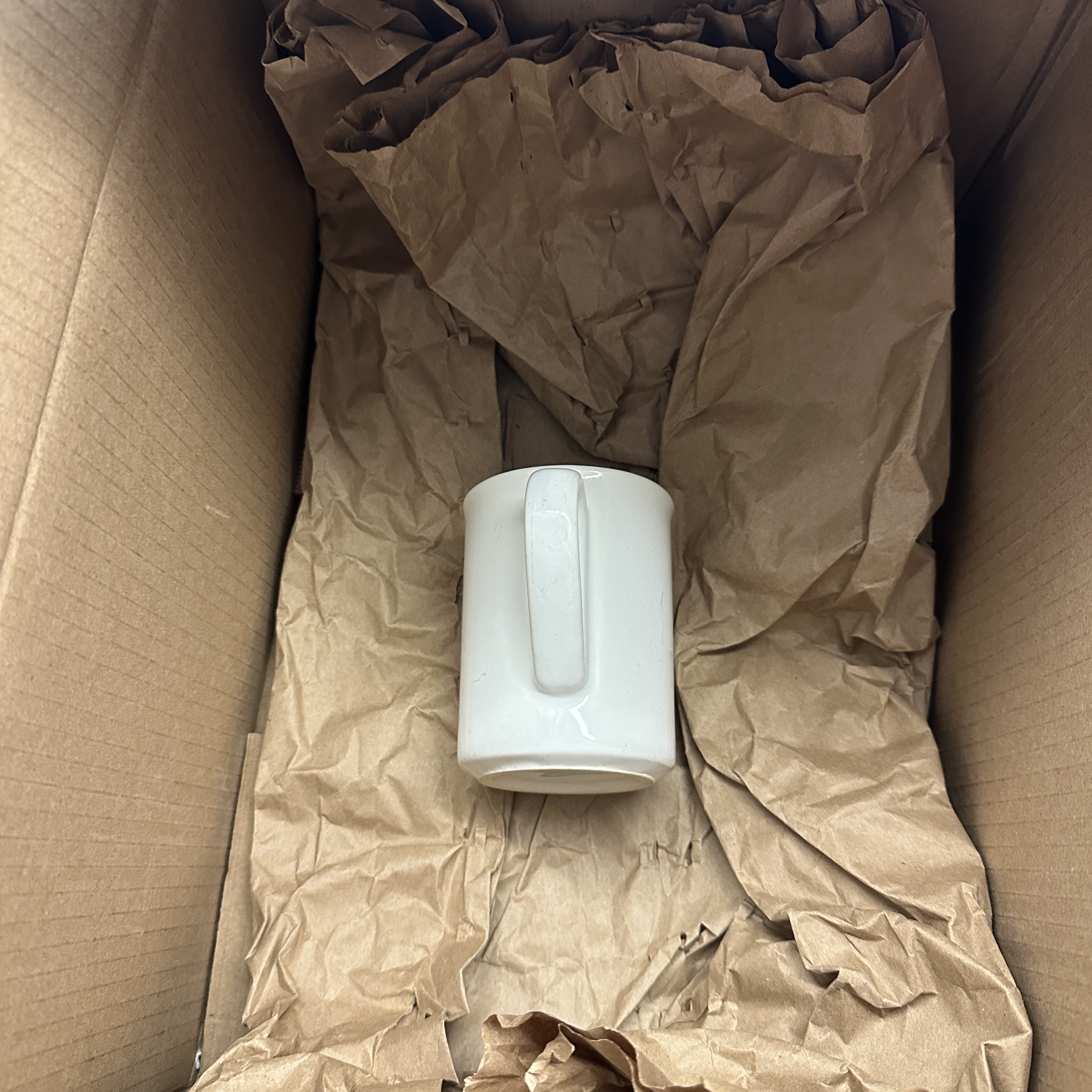}
        \label{fig:setup_b}
    \end{subfigure}
    \hspace{0.5em}
    \begin{subfigure}[b]{0.14\textwidth}
        \centering
        \includegraphics[width=\textwidth]{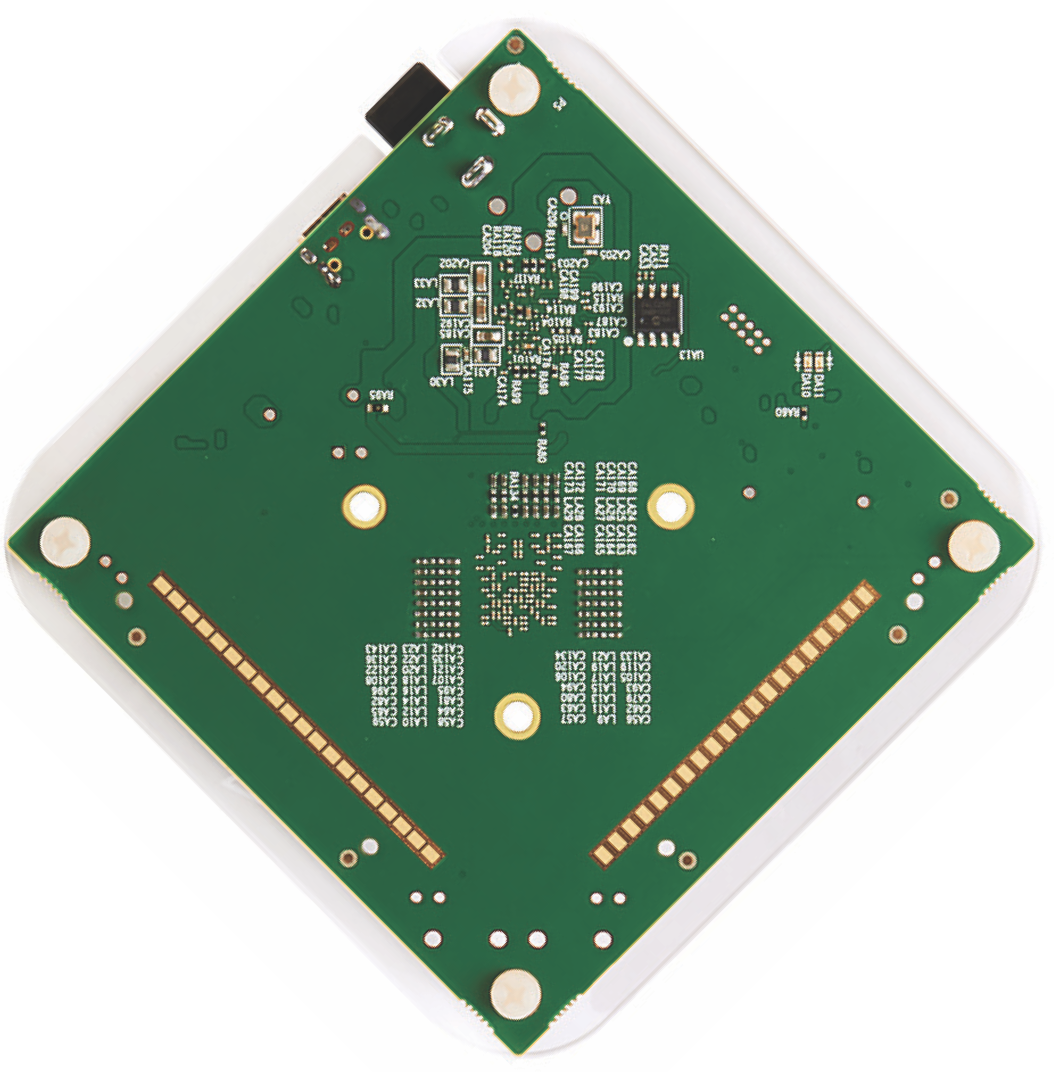}
        \label{fig:setup_c}
    \end{subfigure}
    \caption{Experimental setup for radar data collection, including the measurement environment (left), one example object (middle), and the employed radar sensor (right)~\cite{vayyar}.}
    \label{fig:setup}
\end{figure}

We evaluate the performance of RadarFuseNet against several ablated and baseline models. The evaluation uses IQ radar samples of packaged objects, exploiting the ability of mmWave radar to penetrate the packaging and characterize the objects inside at two different frequency bands. The dataset consists of ten objects, uniformly distributed over all classes: hammer, screwdriver, deodorant, calculator, coiled cable, mug, tape roll, water bottle, plastic cup, and ball.
The data collection setup and the radar sensor are depicted in Fig.~\ref{fig:setup}.
Each sub-dataset (64\,GHz and 67\,GHz) contains 500 samples, resulting in 1000 samples in total. 
A conventional 60/20/20 training, validation, and test split was applied independently to each frequency-specific sub-dataset.
All models are trained with a batch size of 32 using the AdamW optimizer~\cite{adamw}, a weight decay of 0.001, and a learning rate of 0.0001. Training was performed for 40 epochs at 64\,GHz and 80 epochs at 67\,GHz to ensure convergence. To reduce sensitivity to parameter initialization, training was repeated with ten different fixed seeds, and all reported results are averages across these runs.
\begin{table}[htbp]
\centering
\caption{Ablation study on the phase-weighting hyperparameter $\lambda$ for occluded-object classification.}
\label{results-abl}

\begin{tabularx}{\columnwidth}{@{}Xcc@{}}
\toprule
\makecell[l]{\normalfont RadarFuseNet\\\normalfont configuration}
    & Acc. 64\,GHz
    & Acc. 67\,GHz \\
\midrule
$\lambda=0.0$ (magnitude softmax) & 97.60\% & 94.40\% \\
$\lambda=0.1$                  & 97.40\% & 94.70\% \\
$\lambda=0.2$                  & \textbf{97.70\%} & \textbf{94.70\%} \\
$\lambda=0.3$                  & 97.30\% & 94.30\% \\
$\lambda=0.4$                  & 97.20\% & 94.00\% \\
$\lambda=0.5$                  & 97.20\% & 93.50\% \\
$\lambda=0.6$                  & 96.70\% & 94.00\% \\
$\lambda=0.7$                  & 96.70\% & 93.50\% \\
\bottomrule
\end{tabularx}

\end{table}
As shown in Table~\ref{results-abl}, including the phase term improves the overall accuracy in both frequency bands, with the best joint performance achieved at $\lambda=0.2$ with $\lambda=0.1$ being very close. The 64\,GHz band consistently yields slightly higher accuracy than the 67\,GHz band.
\begin{table}[htbp]
\centering
\caption{Overall accuracy of models for occluded-object classification and their ablated versions.}
\label{results-occ}

\setlength{\tabcolsep}{2pt}
\renewcommand{\arraystretch}{1.1}
\footnotesize

\begin{tabularx}{\columnwidth}{@{}Xcc@{}}
\toprule
Model & Acc. 64\,GHz & Acc. 67\,GHz \\
\midrule
Single-branch complex-valued CNN & 94.80\% & 93.50\% \\
Real-valued cross-attention RadarFuseNet & 94.90\% & 91.60\% \\
RadarFuseNet (MLP fusion) & 92.30\% & 87.70\% \\
ACCOR~\cite{accor} & 96.60\% & 93.59\% \\
ResNet-18~\cite{resnet} & 85.70\% & 89.50\% \\
ResNet-34~\cite{resnet} & 82.70\% & 88.10\% \\
\textbf{RadarFuseNet} (ours) & \textbf{97.70\%} & \textbf{94.70\%} \\
\bottomrule
\end{tabularx}
\end{table}
Since FFT representations provide stronger features in distinct range-angle peaks~\cite{smcnet}, all baselines use FFT inputs instead of IQ, while RadarFuseNet fuses both modalities.
Table~\ref{results-occ} shows that our model outperforms all comparison models in both frequency bands, including the single-branch complex-valued CNN, the real-valued RadarFuseNet, the MLP-fusion variant, ACCOR~\cite{accor}, and the ResNet baselines~\cite{resnet}. The single-branch CNN uses only one complex-valued CNN with FFT input, which is evaluated due to its superior performance over raw IQ input~\cite{smcnet}.
The real-valued RadarFuseNet applies standard multi-head cross-attention after concatenating the real and imaginary parts, whereas the MLP-fusion variant replaces cross-attention with two MLP layers applied after the feature extraction.
The ResNet baselines serve as off-the-shelf, real-valued CNN comparisons.
Generally speaking, standard ResNets fail to match or outperform our proposed architecture under the current data and training configurations.
\section{Conclusion and Future Work}
\label{conclusion}
This paper presented RadarFuseNet, a dual-branch architecture that fuses raw radar IQ and FFT representations using complex-valued CNNs and bidirectional phase-weighted complex-valued cross-attention. By incorporating phase alignment into the attention weights, RadarFuseNet effectively combines IQ and FFT features while considering phase information rather than neglecting it. It achieves 97.70\% accuracy on 64\,GHz data and 94.70\% on 67\,GHz data, outperforming all evaluated baselines and ablated variants in radar occluded object classification. Although the limited availability of annotated radar IQ data restricts generalizability, future work will include collecting larger and more diverse datasets for IQ radar-specific tasks, as well as evaluating the proposed fusion mechanism on other fusion tasks involving two complex-valued input representations.

\end{document}